\documentclass[fleqn,10pt]{wlscirep}

\newcommand{\bfpsi}{\mbox{\boldmath$\psi$}}

\title{Competing magnetic ground states and their coupling to the crystal lattice in CuFe$_2$Ge$_2$}

\author[1,*]{Andrew F. May}
\author[2]{Stuart Calder}
\author[1]{David S. Parker}
\author[1]{Brian C. Sales}
\author[1]{Michael A. McGuire}
\affil[1]{Materials Science and Technology Division, Oak Ridge National Laboratory, Oak Ridge, TN 37831, US}
\affil[2]{Quantum Condensed Matter Division, Oak Ridge National Laboratory, Oak Ridge, TN 37831, USA}

\affil[*]{mayaf@ornl.gov}


\begin{abstract}
Identifying and characterizing systems with coupled and competing interactions is central to the development of physical models that can accurately describe and predict emergent behavior in condensed matter systems.  This work demonstrates that the metallic compound CuFe$_2$Ge$_2$ has competing magnetic ground states, which are shown to be strongly coupled to the lattice and easily manipulated using temperature and applied magnetic fields.  Temperature-dependent magnetization $M$ measurements reveal a ferromagnetic-like onset at 228(1)\,K and a broad maximum in $M$ near 180\,K. Powder neutron diffraction confirms antiferromagnetic ordering below $T_{\textrm{N}}\approx$175\,K, and an incommensurate spin density wave is observed below $\approx$125\,K.  Coupled with the small refined moments (0.5-1\,$\mu_B$/Fe), this provides a picture of itinerant magnetism in CuFe$_2$Ge$_2$.  The neutron diffraction data also reveal a coexistence of two magnetic phases that further highlights the near-degeneracy of various magnetic states.  These results demonstrate that the ground state in CuFe$_2$Ge$_2$ can be easily manipulated by external forces, making it of particular interest for doping, pressure, and further theoretical studies.  \href{http://www.nature.com/articles/srep35325}{http://www.nature.com/articles/srep35325}
\end{abstract}
\begin{document}

\flushbottom
\maketitle

\thispagestyle{empty}

\section*{Introduction}

Systems with a strong coupling between magnetism, the crystal lattice, and itinerant electrons often display interesting physics.  When such systems have several nearly-degenerate ground states, complex emergent behavior can be observed, such as unconventional superconductivity.\cite{Dagotto2005}  The behavior of systems with nearly-degenerate ground states can typically be tuned using external forces (pressure, magnetic field) or chemical manipulation, and identifying the so-called parent compounds that are characterized by magnetic and/or structural instabilities is one focus of science-driven synthesis.

CuFe$_2$Ge$_2$ has been predicted to possess an antiferromagnetic (AFM) ground state, with multiple magnetic configurations close in energy\cite{Shanavas2015}. There have been no experimental investigations of its physical properties.  The theoretical calculations revealed multiple bands with a high density of Fe $d$ states at the Fermi level, suggesting an itinerant character of the magnetism.  The associated sheet-like structures of the Fermi surface provide nesting instabilities and thus promote the AFM ground state \cite{Shanavas2015}.  Based on these calculations, Cu is anticipated to be non-magnetic.

\begin{figure}[ht]
\centering
\includegraphics[width=0.9\linewidth]{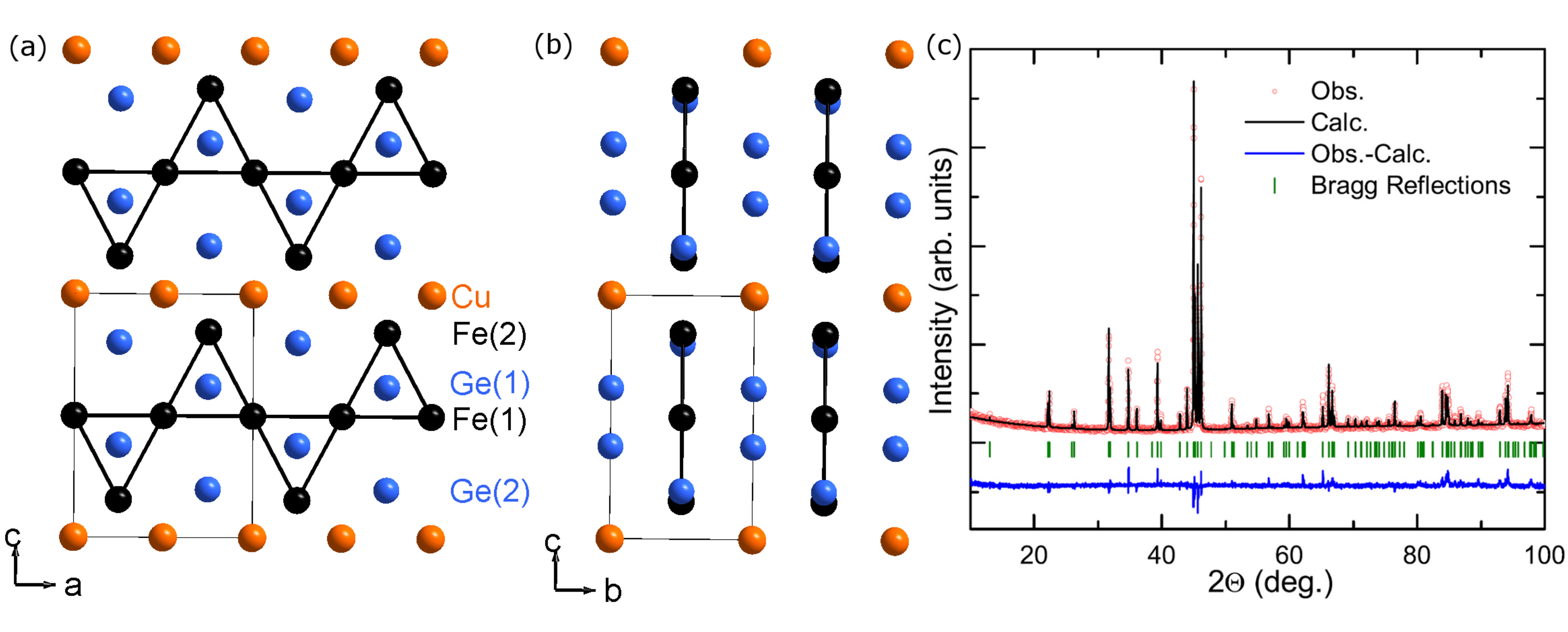}%
\caption{(a,b) Crystal structure of CuFe$_2$Ge$_2$ highlighting the sawtooth chains of Fe atoms residing in the ($ac$) plane.  (c) Powder x-ray diffraction data of CuFe$_2$Ge$_2$ at room temperature.  The calculated curve is from Rietveld refinement, with atomic positions of Cu at 0,0,0; Ge(1) at $\frac{1}{4},0,0.380(1)$; Ge(2) at $\frac{3}{4},\frac{1}{2},0.198(1)$; Fe(1) at $0,\frac{1}{2},\frac{1}{2}$; and Fe(2) at $\frac{1}{4},\frac{1}{2},0.155(1)$.}
\label{structure}
\end{figure}

CuFe$_2$Ge$_2$ is an intermetallic compound possessing an orthorhombic crystal structure (space group 51, \textit{Pmma})\cite{Zavalij1987}.  As shown in Fig.\,\ref{structure}, it is characterized by sawtooth chains of Fe that lie within the $ac$ plane.  Fe(1) forms the centerline of the sawtooth chain running along the $a$-axis; it is coordinated by a distorted octahedra of Ge.  These Fe(1) positions are separated by 2.49\AA\, at room temperature, which is nearly identical to the nearest neighbor in metallic Fe (bcc) at room temperature (2.50\AA)\cite{Raeuchle1954}.  Fe(1)-Fe(2) bonds are $\approx$2.66\AA\,, forming a sawtooth chain of isosceles triangles.  The Fe(2)-Fe(2) distance between sawtooth chains along the $c$-axis is 3.25\AA\,, though bonds with copper link the chains.  Thus, in spite of the three-dimensional crystal structure, the electronic and magnetic properties may be expected to possess significant anisotropy.

This study reports the synthesis and characterization of polycrystalline CuFe$_2$Ge$_2$.  Neutron powder diffraction reveals an incommensurate spin density wave between the base temperature of 4\,K and 125\,K, while commensurate AFM order is observed below the N\'eel temperature of $\approx$175\,K down to 100\,K.  Interestingly, a coexistence of these two magnetic phases is observed in the crossover region.  Magnetization measurements reveal a clear increase in the magnetization below $T_0\approx$228\,K, and a small ferromagnetic component remains upon cooling into the AFM state.  These measurements also demonstrate an evolution of the magnetization as a function of applied field, with relatively small changes in the field strongly affecting the observed behavior.  In addition, the magnetism is found to be strongly coupled to the lattice, as demonstrated by anisotropic thermal expansion and a strong change in the $c$ lattice parameter at the magnetic phase transitions.  In total, these results demonstrate that CuFe$_2$Ge$_2$ has complex magnetism resulting from nearly-degenerate states that can be easily manipulated.  The application of external forces or chemical doping will likely produce additional emergent states in CuFe$_2$Ge$_2$, potentially including superconductivity.

\section*{Results}

\subsection*{Zero-Field Limit of Magnetism}

\begin{figure}[ht!]
\centering
\includegraphics[width=0.9\linewidth]{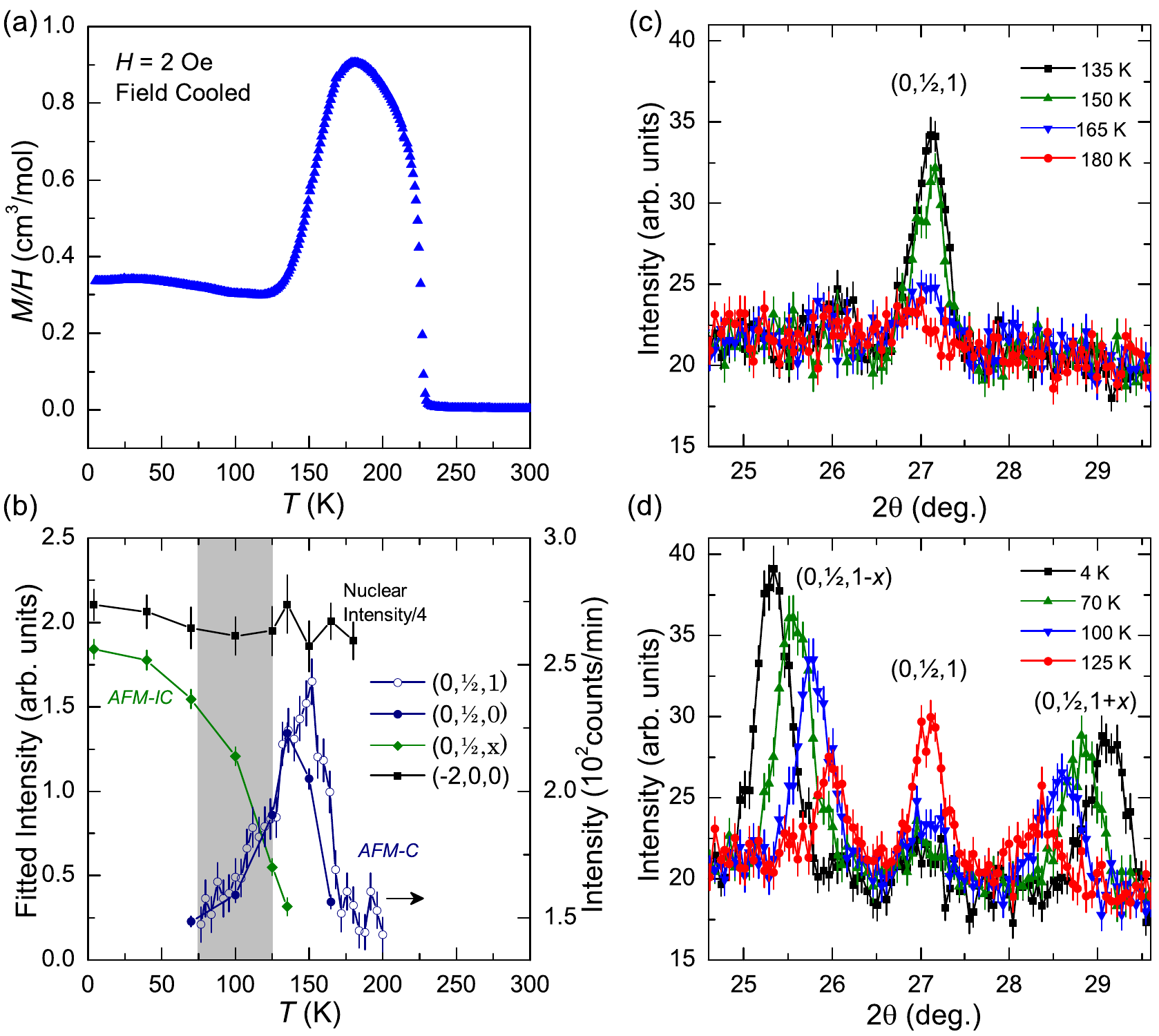}%
\caption{Magnetic phase transitions in CuFe$_2$Ge$_2$ illustrated using (a) temperature-dependent magnetization and (b-d) neutron diffraction data. (b) Intensity at magnetic and nuclear Bragg reflections as a function of temperature; data for the (0,$\frac{1}{2}$,1) reflection were collected using a fixed $|$Q$|$=1.21\AA\,= 26.9\,degrees 2$\theta$, whereas the other intensities come from peak fitting. (c) Magnetic Bragg reflection with (0,$\frac{1}{2}$,0) propagation vector and (d) emergence of additional magnetic reflections at 125\,K, which come from an incommensurate spin structure with propagation vector (0,$\frac{1}{2}$,$x$).  The magnetic Bragg reflections are indexed.}
\label{MT}
\end{figure}

Figure \,\ref{MT} provides an overview of the temperature-dependent data utilized to characterize the magnetic phase transitions in CuFe$_2$Ge$_2$.  The temperature dependence of the magnetization $M$ was found to depend strongly on the applied field $H$.  To probe the magnetic ground state, data were collected while cooling in a very small field of $H$=2\,Oe; the data are plotted as $M/H$, which equals the magnetic susceptibility $\chi$ when $M$ is linear with $H$.  As observed in Fig.\,\ref{MT}a, $M$ increases sharply upon cooling below $T_0\approx$228\,K, which appears similar to the onset of ferromagnetic order.  However, the magnetization reaches a maximum near 181\,K and then decreases upon cooling until essentially plateauing below $\approx$\,120\,K.  The decrease in $M$ upon cooling suggests antiferromagnetic (AFM) order, though the shape of $M(T)$ is unusual and it is difficult to define a N\'eel temperature $T_N$ from these data.  Therefore, we define the maximum in $M(T)$ as $T^*$, and obtain $T_N$ from neutron diffraction data that clearly demonstrate AFM order.  The shape of $M(T)$ indicates unusual magnetism or competition between different magnetic states. The magnetic susceptibility has a large temperature-independent contribution above $\approx$300\,K, and high $T$ susceptibility measurements are necessary to further probe the potential role of localized moments in the non-magnetically-ordered state.

Neutron diffraction data demonstrate AFM order below $T_N\approx$175\,K, as well as the transition from a commensurate ($AFM-C$) to incommensurate ($AFM-IC$) spin structure upon cooling below $\approx$125\,K.  These transitions are shown in Fig.\,\ref{MT}b, where the intensities of several neutron diffraction reflections are plotted as a function of temperature.  AFM order is demonstrated by intensity at the (0,$\frac{1}{2}$,0) and (0,$\frac{1}{2}$,1) Bragg reflections.  This magnetic scattering is described by the propagation vector \textbf{k}$_{C}$=(0,$\frac{1}{2}$,0), which indicates the moments are antiferromagnetically-aligned along the $b$-axis.  The intensity of magnetic Bragg peaks from $AFM-C$ starts to increase near 175(5)\,K, reaches a maximum near 135\,K, and then decreases to near background levels at the base temperature of 4\,K.  At 125\,K, a set of magnetic Bragg reflections from $AFM-IC$ are observed (such as those at 26 and 28.4 \,deg.\,2$\theta$ in Fig.\,\ref{MT}d).  These magnetic reflections are indexed to the incommensurate propagation vector  \textbf{k}$_{IC}$=(0,$\frac{1}{2}$,$x$), and their intensity increases upon cooling to the base temperature of 4\,K.  In Fig.\,\ref{MT}d, the two outermost peaks move apart upon cooling, which corresponds to an increase in $x$ (a decrease in the real-space sinusoidal period).  Refinement yields $x$=0.07765 (125\,K), 0.08723 (100\,K), 0.1044 (70\,K) and 0.117 (4\,K).  A clear coexistence of $AFM-C$ and $AFM-IC$ is observed between approximately 70 and 125\,K, indicating a competition between magnetic phases in this region.  Note that at 200\,K neutron diffraction intensity is not detected at locations of the peaks now identified as magnetic contributions (see Supplementary Information), and x-ray diffraction does not reveal intensity at these locations.

These results demonstrate the competition between two or more magnetic structures in CuFe$_2$Ge$_2$, and the transition between a commensurate and an incommensurate magnetic state may have implications regarding the movement of electronic bands as a function of temperature.  In elemental chromium, the prototype itinerant AFM, the spin density wave is incommensurate and a spin-flop is observed at about 0.4$T_N$.  The SDW in chromium is driven by electronic nesting, and an imbalance in electron and hole pockets leads to the incommensurate structure.  Due to the electronic origin, the SDW can be easily manipulated by doping.  For instance, the addition of Mn leads to the stabilization of commensurate magnetic order, and for some small concentrations of Mn a transition between commensurate and incommensurate SDWs is observed upon cooling\cite{Geerken1982}.  These transitions in Cr-Mn have been considered within the framework of defects and electronic damping\cite{Fishman1993}.  It would be interesting to see if the transition between $AFM-C$ and $AFM-IC$ in CuFe$_2$Ge$_2$ were influenced by anti-site defects or other intrinsic dopants. Such chemical disorder may be difficult to identify in CuFe$_2$Ge$_2$ due to the similar atomic masses.  However, the  changes with annealing $T$ that are discussed in the Supplementary Information may allow for future experiments that examine the influence of disorder on the observed transitions.

\begin{figure}[t]
\centering
\includegraphics[width=0.9\linewidth]{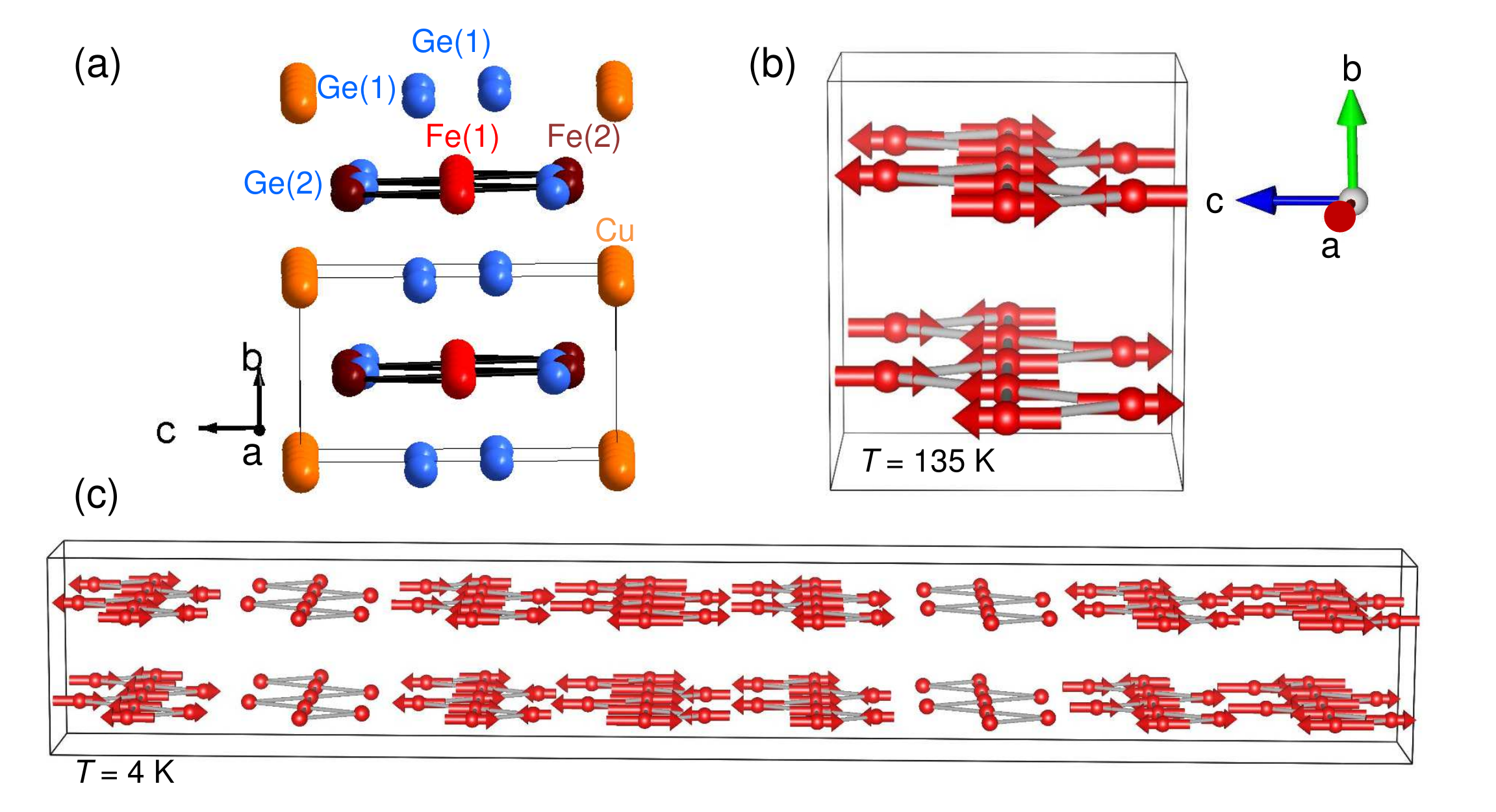}%
\caption{Comparison of nuclear and magnetic structures in CuFe$_2$Ge$_2$.  (a) Two units cells of the nuclear structure, with atomic positions labeled. (b,c) The magnetic models utilized at (b) 135\,K and (c) 4\,K, with only the Fe positions and moments shown.  The spin structure shown in (c) is a commensurate structure with propagation vector (0,$\frac{1}{2}$,$\frac{1}{8}$), which closely represents the observed incommensurate structure with propagation vector (0,$\frac{1}{2}$,0.117) at 4\,K.}
\label{MagStructure}
\end{figure}

At 135\,K, the neutron data can be fit to a spin structure that is antiferromagnetic along $b$, with chains of Fe(1) atoms ferromagnetically coupled  along $a$ and antiferromagnetically coupled with Fe(2).  This  corresponds to the irreducible representation $\Gamma$4 (see Methods).  The magnetic structure obtained for the $AFM-C$ phase is consistent with the `\textit{AF-G}' ground state predicted by Shanavas and Singh\cite{Shanavas2015}.  Unfortunately, the neutron data can be described equally-well when the antiferromagnetically-aligned Fe moments point along $a$ or along $c$ (or a combination of the two).  Our first principles calculations find that the $c$-axis is the preferred orientation of the moments (see Methods for details), and thus the Fe moments were constrained to point along $c$.  The corresponding magnetic structures are shown in Fig.\,\ref{MagStructure}, and we emphasize that these are the simplest models that describe the data.  The refined moments are 0.36(10)\,$\mu_B$/Fe(1) and 0.55(10)\,$\mu_B$/Fe(2) at 135\,K.  Small moments can be expected near $T_N$, though this observation also lends support to the hypothesis that the magnetism is itinerant in CuFe$_2$Ge$_2$.  Note that the system is metallic in regards to the temperature-dependence of its electrical resistivity (see Supplementary Information).

Given the theoretical prediction of itinerant magnetism, and the experimentally confirmed reduced magnetic moment, the incommensurate phase $AFM-IC$ is treated with a spin density wave (SDW) model, though a helical structure would also describe the scattering.  The proposed structure for $AFM-IC$ shown in the lower portion of Fig.\,\ref{MagStructure} is represented by a commensurate structure with propagation vector (0,$\frac{1}{2}$,$\frac{1}{8}$); the measured propagation vector is (0,$\frac{1}{2}$,0.117) at 4\,K.  Refinement of the 4\,K data yields moments of 1.0(1)\,$\mu_B$/Fe(1) and 0.71(10)\,$\mu_B$/Fe(2).  These represent the maximum moments because they vary as a function of $z$.  Our first principles calculations (using LDA) obtained $\pm 1.25\,\mu_{B}$/Fe(1) and $\pm 0.90\,\mu_{B}$/Fe(2) for the magnetic ground state AF0 (see Methods); note that the symmetry of AF0 is consistent with $AFM-C$.  Reasonable agreement between theory and experiment is observed, especially considering that the calculation is being performed in a different magnetic symmetry than the experimentally-observed incommensurate structure.  For instance, the agreement here is better than that in the Fe-based superconductors, where theory significantly overestimated the Fe moments despite utilizing the experimentally-observed magnetic structures.\cite{SinghReview2009}

\subsection*{Coupling of Magnetism and the Crystalline Lattice}

Shanavas and Singh reported first principles calculations that demonstrated various magnetic symmetries/configurations are within 20\,meV of each other\cite{Shanavas2015}.  Our calculations are consistent with this, and thus the observed properties may be understood as a consequence of nearly-degenerate ground states.  However, the presence of nearly-degenerate ground states does not necessarily explain the evolution of magnetism, especially in a system that orders at fairly high $T$.  Therefore, an additional energy or order parameter must be important in driving the magnetic transitions.

\begin{figure}[ht]
\centering
\includegraphics[width=\linewidth]{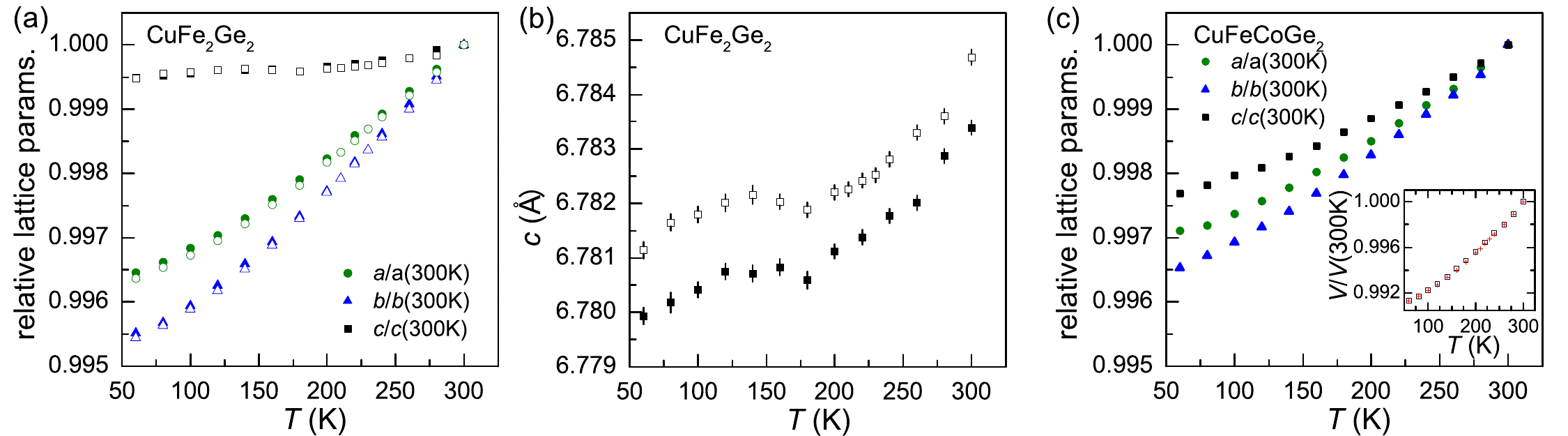}%
\caption{(a) Evolution of relative lattice parameters as a function of $T$ in CuFe$_2$Ge$_2$. (b) The refined $c$ lattice parameter in CuFe$_2$Ge$_2$.  Data for two samples are shown, with the open symbols representing the sample that was also used for neutron diffraction.  (c) Relative lattice parameters versus $T$ in the doped, non-magnetically-ordered composition CuFeCoGe$_2$ demonstrating a smooth contraction with decreasing $T$ in the absence of magnetic order. The inset shows the relative change in unit cell volumes, with red crosses for CuFe$_2$Ge$_2$ and open squares for CuFeCoGe$_2$. In (a,c) the error bars are smaller than the data markers.}
\label{lattice}
\end{figure}

One driving force for stabilizing a particular magnetic phase can be magnetoelastic coupling.  To probe this behavior, x-ray diffraction data were collected as a function of temperature (see Fig.\,\ref{lattice}).  Consideration of magnetoelastic effects is greatly aided by comparison with a non-magnetic analogue.  For this system, we found that a sample of nominal composition CuFeCoGe$_2$ provides a non-magnetically-ordered baseline to consider.  Magnetization data show that this composition appears paramagnetic down to at least 5\,K, though a small anomaly was observed near 150\,K (see Supplementary Information for powder x-ray diffraction and magnetization data). The x-ray diffraction data for CuFeCoGe$_2$ are well-described using the CuFe$_2$Ge$_2$ model, and at room temperature the lattice parameters are $a$=4.9966(1)\AA\,, $b$=3.9262(1)\AA\,, and $c$=6.7162(1)\AA. Compared to $a$ = 4.9765(1)\AA, $b$ = 3.9718(1)\AA, and $c$ = 6.7834(1)\AA\ for CuFe$_2$Ge$_2$ at 300\,K, these results show that cobalt incorporation increases $a$ while $b$ and $c$ decrease.  The unit cell volume of CuFeCoGe$_2$ is nearly 2\% smaller than that of CuFe$_2$Ge$_2$.

The thermal expansion is found to be anisotropic in CuFe$_2$Ge$_2$, with $c$ changing much less than $a$ or $b$ (Fig.\,\ref{lattice}a).  As shown in Fig.\,\ref{lattice}b, the $c$ lattice parameter has a non-monotonic $T$-dependence and responds strongly to the magnetic order near $T_N\approx175$\,K and again below $T\approx$120\,K (where the incommensurate structure begins to dominate).  This is not the case for CuFeCoGe$_2$, where the thermal expansion is much more isotropic and monotonic (Fig.\,\ref{lattice}c).  As seen in the inset of Fig.\,\ref{lattice}c, the relative changes in the unit cell volumes are nearly identical for the doped and undoped samples.  This shows that expansion along $a$ and $b$ are enhanced for CuFe$_2$Ge$_2$ in response to the stiff behavior along $c$.  It is worth noting that the sawtooth chains of Fe exist in the $ac$ plane, running along $a$ with the shortest distance between chains being along $c$.  These results clearly demonstrate a strong coupling between the lattice and the magnetism in CuFe$_2$Ge$_2$, and in particular they suggest that the application of anisotropic strain would likely have a significant impact on the ground state in CuFe$_2$Ge$_2$.

First principles calculations were performed to further investigate the role of magnetoelastic coupling in CuFe$_2$Ge$_2$.  An overview of the results is provided here and additional details are given in the Supplementary Information.  In the parent phases of the Fe-based superconducting (SC) systems, where strong magnetoelastic coupling is observed, the calculated moments were found to vary strongly with the Fe-As bond length.\cite{Yin2008,Belashchenko2008,SinghReview2009,JohnstonReview}  To check for similar behavior in CuFe$_2$Ge$_2$, the moments were calculated as a function of the $z$ coordinate of Fe(2).  As $z$ increases from its equilibrium position, moving Fe(2) toward Fe(1), the moment on Fe(1) increases slightly and the moment on Fe(2) decreases.  A relative increase in the distance between Fe(1) and Fe(2) causes very little change in the moments.  These different dependencies on $z$ illustrate a difference between CuFe$_2$Ge$_2$ and the Fe-based superconductors, and are one manifestation of the relatively complex magnetic interactions and competing ground states in CuFe$_2$Ge$_2$.  In a complementary calculation, the influence of the imposed magnetic symmetry on this $z$ coordinate was investigated by relaxing the atomic positions for different magnetic states.  These calculations found that $z$ increases by 7.5\% in the ferromagnetic state relative to the AFM ground state (bringing Fe(1) and Fe(2) closer together).  The $z$ of Fe(2) only changed by 1.3\% between the AFM ground state and the non-magnetic state, though.  Refinement results from x-ray diffraction data did not reveal any significant changes in atomic positions as a function of $T$ as shown in the Supplementary Information.

\subsection*{Field Dependence of Magnetism}

\begin{figure}[ht]
\centering
\includegraphics[width=\linewidth]{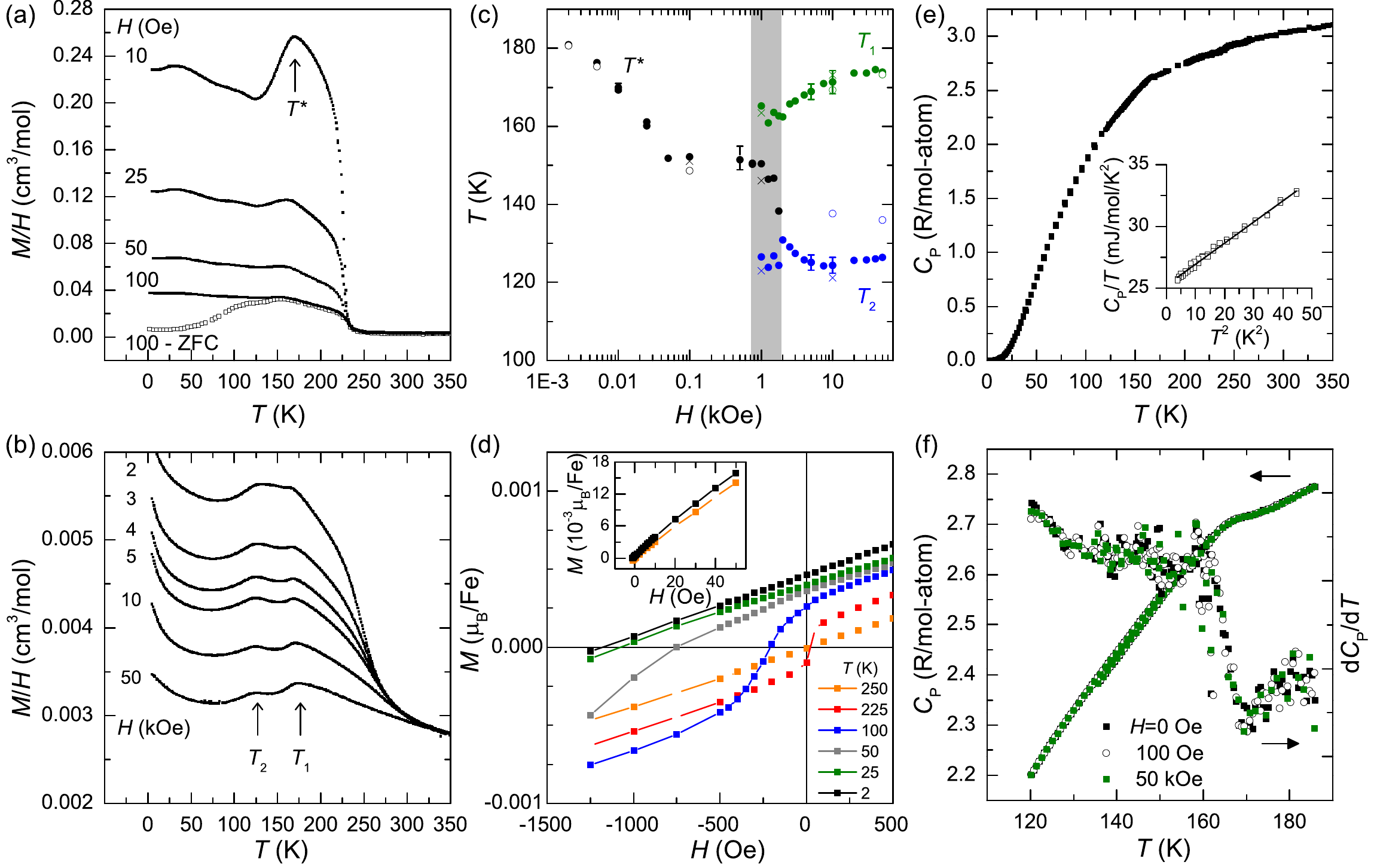}%
\caption{Influence of applied magnetic field on the temperature dependence of magnetization and specific heat.  (a) $M(T)/H$ is shown for small fields where a single maximum at $T^*$ is observed, and zero-field cooled data (ZFC) are shown for 100 Oe.  (b) The evolution of $M(T)$ with applied fields revealing critical temperatures $T_1$ and $T_2$ for $H\,\gtrsim$\,2\,kOe. (c) Phase diagram of the critical temperatures from $M(T)$ curves, with error bars shown for a few points.  These temperatures are poorly defined in the shaded region where all three are observed. Data for multiple samples are shown for comparison. (d) Isothermal magnetization curves at various temperatures highlighting the low-field data that demonstrate an increase in the small ferromagnetic component with decreasing $T$; the inset shows $M(H)$ is mostly linear at 2 and 250\,K. (e) Specific heat capacity data with low-$T$ fit shown in the inset. (f) The anomaly in the specific heat capacity near 170\,K is found to be unchanged by applied fields of 100\,Oe and 50\,kOe; the derivative is plotted on the right-side axis.}
\label{field}
\end{figure}

The application of a magnetic field changes the shape of $M(T)$, as shown in Fig.\,\ref{field}(a,b).  In Fig\,\ref{field}a, a single maximum defined as $T^*$ can be observed, but the application of fields $H>2$\,kOe leads to the presence of two distinct maximum at $T_1$ and $T_2$ (see Fig\,\ref{field}b).  This is a relatively small field to produce such a large change in the magnetism at these temperatures.  Specifically, because the moments are small the total magnetic energy is small; for a moment of 1$\mu_B$ the magnetic energy is equivalent to 0.67\,K for every 10\,kOe applied.  As such, it is surprising that the magnetism can be changed by such a small field and these results may also have implications for the response of the lattice to an applied field.

The phase diagram demonstrating the evolution of $T^*$, $T_1$, and $T_2$ with applied field is shown in Fig.\,\ref{field}c.  $T^*$ decreases abruptly with applied field; similar behavior is typically observed for the N\'eel temperature of an antiferromagnetic transition, although usually with a much weaker field dependence.  A region between approximately 700 and 2000\,Oe appears to have a coexistence of all three characteristic temperatures, and it is difficult to define $T^*$, $T_1$, and $T_2$ in this range (though estimates are shown).  In addition, the $M(T)$ curves have rather broad features/maxima, making the definitions of critical temperatures especially difficult for intermediate fields.  Outside of this region, the typical error for these critical temperatures is about 3\,K for a given sample.  Data for three different samples are included in Fig.\,\ref{field}c (represented by closed circles, open circles, and $\times$).   The different samples possessed similar behavior, though a notable difference for $T_2$ was observed in the sample used for neutron diffraction (open circles).  This may suggest that sample possessed a different amount of atomic disorder or interstitial occupancy.  Importantly, however, the behavior of $T^*$ is consistent among all samples, indicating that the low-field behavior of all samples is consistent and thus the neutron diffraction results should be representative of this ground state.  While not shown, $T_0$ associated with the onset of magnetization increases with increasing $H$, as expected for ferromagnetic order.

Isothermal magnetization measurements at various temperatures are shown in Fig.\,\ref{field}d, where the main panel emphasizes the small ferromagnetic contribution observed at low fields.  As shown in the inset, the $M(H)$ curves are nearly linear for all temperatures, consistent with a paramagnetic state or antiferromagnetic order.  A small ferromagnetic component is observed starting at 225\,K, consistent with the sharp rise in magnetization near $T_0\approx$228\,K; the remanent moment and coercivity increase with decreasing $T$.  Since neutron diffraction clearly shows this to be antiferromagnetically ordered below $\approx$175\,K, we speculate that the ferromagnetic component (the remanence, coercivity) arises due to either a canting of the antiferromagnetically-aligned moments or from the magnetization of interstitial or anti-site Fe species.  The FM component increases as the annealing temperature increases (and crystallinity decreases), but the temperature-dependence remains the same.  This suggests that the degree of canting or the number of FM sites is dependent on the concentration of point defects (site disorder, vacancies, interstitials).  As such, it does not appear that the FM component originates from an impurity phase.

Specific heat capacity data are shown in Figure \ref{field}e, and Fig.\,\ref{field}f highlights data collected near $T_N$. The anomaly at $T_N$ is subtle, with an onset near 170\,K as shown by the derivative curve in Fig.\,\ref{field}f.  Interestingly, the $C_P$ anomaly does not change in applied fields, which is fairly unusual for magnetic phase transitions and is especially surprising for one that is easily influenced by an applied field.  The lack of an anomaly in $C_P$ near $T_2\approx$125\,K for $H$=50\,kOe suggests that the feature in $M(T)$ at $T_2$ is related to a reconfiguration of the moments.  That is, if the magnetic entropy is quenched at $T_N$ or $T_1$ then the anomaly in $M(T)$ at $T_2$ should not have a corresponding anomaly in $C_P$.   Therefore, $T_2$ may be related to a cross-over between the commensurate and incommensurate antiferromagnetic structures, or perhaps it is associated with a reorientation of the moments (between crystallographic axes).  In either case, the emergence of $T_2$ clearly demonstrates the sensitivity of magnetism in CuFe$_2$Ge$_2$ to applied fields.

The inset of Fig.\,\ref{field}e highlights the low $T$ data, which were fit to $C_P/T$ = $\gamma$ + $\beta T^2$, where $\gamma$ is the Sommerfeld coefficient associated with electronic contributions and $\beta$ represents the Debye phonon contribution; note that we are neglecting magnetic contributions.  A relatively large Sommerfeld coefficient of $\gamma$=25.2\,mJ/mol/K$^2$ was obtained; from $\beta$ a Debye temperature of 385\,K was calculated.  The large value of $\gamma$ is likely associated with the presence of Fe $d$ states near the Fermi level.  Using first principles calculations, Shanavas and Singh obtained a bare (non-magnetically ordered) Sommerfeld coefficient of 18.6\,mJ/mol/K$^2$, suggesting that the Fermi surface naturally has a large density of states that promotes itinerant magnetism\cite{Shanavas2015}.  For comparison, $\gamma$=16\,mJ/mol/K$^2$ in BaFe$_2$As$_2$\cite{Rotter2008}, which is a parent compound for Fe-based superconductors and has a spin density wave transition near 140\,K.

\section*{Discussion and Summary}

The data reveal that CuFe$_2$Ge$_2$ has competing magnetic states with dominant antiferromagnetic order at low $T$.  Neutron diffraction demonstrates AFM order below $\approx$175\,K, and an incommensurate spin structure is observed below $\approx$125\,K.  The magnetization has an unusual temperature dependence with a broad maximum near $T_N$, and the results suggest proximity to ferromagnetic ordering.  A weak ferromagnetic contribution is evident in the magnetization data below $\approx$\,228\,K and divergence of field-cooled and zero-field-cooled data is also observed.  Comparison of the zero-/low-field results to the field-dependent data provides some insights into the nature of the magnetic ordering.  For instance, the $C_P$ anomaly occurs roughly 9\,K below the maximum of $M(T)$ observed for an applied field of 2\,Oe (the largest $T^*$ in Fig.\,\ref{field}d).  The $C_P$ anomaly does not change with applied field, while the maximum in $M(T)$ is strongly suppressed with increasing $H$.  These discrepancies demonstrate the difficulty in defining $T_N$ based on the $M(T)$ data due to the FM-like onset and the unusual shape of $M(T)$.  In a more general sense, though, these results may point to a gradual and complex evolution of the magnetism.  For instance, the onset of magnetization may be associated with local ordering or canted-antiferromagnetic ordering of one Fe sublattice; a complex ferrimagnetic arrangement may also exist. Local probe measurements, such as M\"ossbauer spectroscopy, may prove particularly insightful.  Measurements on single crystals would certainly prove useful.  However, due to the decomposition/disordering of the phase above 630$^{\circ}$C, care must be taken during crystal growth.  In this regard, a measurement of the magnetization appears to be a useful probe of sample quality; smaller $M$ and smaller ferromagnetic-contributions at low $T$ seem to indicate higher sample quality.

In summary, CuFe$_2$Ge$_2$ possesses complex magnetism with a magnetic structure that evolves as a function of temperature and applied field.  In addition, cobalt doping has been shown to suppress the magnetic order.  Experimental and theoretical results suggest magneto-structural effects play an important role in determining the magnetic ground state.  These results also clearly demonstrate that the magnetism in CuFe$_2$Ge$_2$ has a strong itinerant character, as illustrated by neutron diffraction data revealing small Fe moments and an incommensurate spin density wave below $\approx$125\,K.  Unlike the Fe-based superconductors, a competition between various magnetic ground states leads to a complex magnetic structure and an unusual temperature-dependence of the magnetization.  Given these observations, studies into the pressure-dependence and doping-dependence of the physical properties in CuFe$_2$Ge$_2$ will likely demonstrate the ability to tune between different magnetic states, and could potentially result in other emergent behavior such as superconductivity.

\section*{Methods}

\subsection*{Synthesis and Characterization}

Polycrystalline samples of CuFe$_2$Ge$_2$ were prepared by arc melting high-purity elements on a cooled-copper hearth, followed by grinding, cold-pressing and annealing.  The annealing temperature was found to be critical, and the samples reported upon here were annealed at 600$^{\circ}$C for 7\,d.     As discussed in the Supplementary Information, significant degradation of the crystallinity is observed when the samples are annealed at 700$^{\circ}$C as opposed to 600$^{\circ}$C.  This is likely due to decomposition and/or disordering of the phase above approximately 615-635\,$^{\circ}$C.

Phase purity was investigated using a Panalytical X'PERT PRO diffractometer with monochromatic copper K$_{\alpha,1}$ radiation, and temperature-dependent structural information was obtained by employing an Oxford closed cycle cryostat.  A representative Rietveld refinement is shown in Fig.\,\ref{structure}c.  The samples were observed to be phase pure to within $\approx$1\%, which is roughly the limit of laboratory x-ray diffraction.  Rietveld refinement using the published orthorhombic structure yielded $a$ = 4.9765(1)\AA, $b$ = 3.9718(1)\AA, and $c$ = 6.7834(1)\AA\ at 300\,K.  For the data in Fig.\,\ref{structure}c, the standard refinement quantifiers\cite{FullProf} are $R_p$=3.38, $R_{wp}$= 4.29, and $\chi^2$=1.36.

Physical property measurements were completed in Quantum Design systems (Physical and Magnetic Property Measurement systems) using standard practices for sample mounting and contacting. Ag epoxy was employed for contacts during four point electrical and thermal transport measurements, and N-grease and H-grease were used for specific heat measurements at $T<200$ and $T>200$\,K, respectively.

\subsection*{Neutron Diffraction Data Collection and Analysis}

Neutron diffraction data were collected on the powder neutron diffractometer HB2A at the High Flux Isotope Reactor in Oak Ridge National Laboratory.  A sample of $\approx$ 8\,g was held in an Al can and data were obtained using a neutron wavelength of 2.4123\AA\,.  This set-up is optimized for measuring small magnetic peaks, but limits the structural refinement details.  Neutron and x-ray diffraction data were refined using the program FullProf\cite{FullProf} and the representational analysis was performed with SARAh\cite{SARAh}.

Representational group analysis was performed to obtain symmetry-allowed magnetic structures from the neutron powder diffraction data at 135\,K and 4\,K (see Supplementary Information). This yields three common irreducible representations (IRs) for the Fe sites ($\Gamma_{mag}$ = $\Gamma_{2}$ + $\Gamma_{4}$ + $\Gamma_{8}$). The observation of a (0,$\frac{1}{2}$,0) reflection implies that the magnetic moment has a component perpendicular to the $b$ axis and this therefore suggests the IR description composed of $b$-axis parallel spins can be discarded ($\Gamma_{2}$).  Representational analysis limits the number of symmetry allowed magnetic structures, however within these the various combinations of couplings between the different Fe sites (basis vectors) necessitates further information to obtain a robust magnetic structure. For these data, it is not possible with the limited number of peaks to distinguish between the IRs $\Gamma_{4}$ and $\Gamma_{8}$.  As such, first principles calculations were utilized to provide insight into the preferred orientation of the magnetic moments.

It is very possible that, particularly for the incommensurate SDW, the orientation of the moments is not along a high-symmetry direction.  The data have been refined using a modulated spin structure (sinusoidal modulation of the moment), which produces zero moment on some Fe sites.  Note that the magnetic scattering can also be described using a helical spin structure and polarized measurements on single crystals would be required to differentiate between these two types of incommensurate structures.  Given the apparent itinerant nature of this magnetism, a SDW model seems more appropriate than a helical one.  The data above 130\,K can be refined using the $AFM-IC$ structure with very small $x$.  However, the simplest magnetic structure that describes the data was utilized, yielding the magnetic structure in Fig.\,\ref{MagStructure}a.  The coexistence of the two magnetic phases at 125\,K may also suggest that the structures at 4\,K and 135\,K are fundamentally different, as one may expect a continuous evolution of $x$ with $T$ if the structure at 135\,K were a nearly-commensurate limit of the incommensurate structure (meaning there would not be a region of coexistence).

Complete neutron diffraction data sets were collected at several temperatures to investigate the nature of the observed magnetic transitions.  These data and corresponding Rietveld refinements are shown in the Supplementary Information.  At 200\,K, the neutron diffraction data are well described by a non-magnetic model that is consistent with the structure used for x-ray diffraction data.  Note that the magnetization data suggest a ferromagnetic-like onset at 228\,K.  Below $T_N$, the observed magnetic scattering is not particularly strong relative to the nuclear Bragg peaks, and thus it is beyond the limits of the measurement to observe a small ferromagnetic component in this material.

\subsection*{First Principles Calculations}

The electronic structure calculations were performed using the all-electron planewave code WIEN2K \cite{wien}.  All results presented in this paper were performed using the experimental lattice parameters, with internal coordinates not dictated by symmetry optimized until the ionic forces were less than 2 mRyd/Bohr. The calculations of magnetostructural properties employed the generalized gradient approximation (GGA) of Perdew, Burke and Ernzerhof \cite{dft:pbe} while the calculations of easy axis orientation used the local density approximation (LDA).  For the GGA calculations an $RK_{max}$ of 7.0 and the augmented plane wave basis was used, where RK$_{max}$ is the product of the smallest LAPW sphere radius and the largest planewave expansion vector.  For the LDA calculations an $RK_{max}$ of 8.0, and the linearized augmented plane-wave basis, were used.  For all calculations, sphere radii of 2.09 Bohr for Ge, 2.20 for Fe and 2.24 for Cu were employed. For all self-consistent calculations, approximately 1000 $k$-points in the full Brillouin zone were used.  It is difficult to perform first principles calculations on non-collinear magnetic structures, and thus we have limited our analysis to collinear configurations.
 
The magnetic ground state is termed AF0 and is consistent with the experimental data at 135\,K; AF0 is equivalent to the \textit{AF-G} ground state found in the previous work \cite{Shanavas2015}.  The first excited state is termed AF1. In both antiferromagnetic states the Fe(1)-Fe(1) nearest neighbor (nn) chains (along the a-axis) are aligned ferromagnetically and the next nearest-neighbor (nnn) Fe(2) is anti-aligned to Fe(1); in AF1 the nnnn (along the $b$-axis) Fe(1) is aligned to the base cell Fe(1) and in AF0 this nnnn Fe1 is antialigned.  We could not stabilize an $ab$-plane checkerboard state, as was studied in the previous work; calculations initialized in this pattern converged to the AF0 state.

Magnetoelastic effects were probed using AF0, AF1, a non-magnetic and a ferromagnetic configuration. The experimental lattice parameters at $T$=220\,K were imposed, and then the internal coordinates that are not dictated by symmetry were optimized.  The greatest change was observed for $z$ of Fe(2), and the obtained values are placed in the Supplementary Information along with the calculated energies.   The non-magnetic state has an optimized $z$ for Fe(2) of 0.1541, and the AFM ground state has a similar $z$ of 0.151.  This is in fair agreement with the experimentally-obtained $z$ of 0.155(1) between 60 and 300\,K.  In addition, the moments within AF0 were calculated as a function of the Fe(2) $z$ after fixing all other crystallographic parameters (coordinates for Ge(1) and Ge(2) were optimized first).  A plot summarizing these results is shown in the Supplementary Information.

In order to guide the refinement of the neutron diffraction data, very careful calculations of the total energy for Fe moments oriented along the three different axes were conducted within the AF0 ground state with the incorporation of spin-orbit coupling.  Note that the magnetic pattern itself - the relative orientation of moments on different Fe sites - is unchanged for these calculations. What we study here is the {\it coupling} of the moments to the crystalline structure, or effectively the magnetic anisotropy. As is well known, this anisotropy derives from spin-orbit coupling, which is expected to be weak here as all elements are relatively light (one recalls the effective $Z^{4}$ dependence \cite{parker_UMn2Ge2} of spin-orbit coupling).  Hence these energy differences are expected to be very small. Thus, a minimum of 10,000 $k$-points were used for these total energy calculations.  Increasing the number of $k$-points to 30,000 changed energy differences by less than 1\%.  These calculations were done using the experimental lattice parameters at 60\,K and 220\,K, and little difference was observed; the 60\,K results are reported.   The moment orientation parallel (and anti-parallel) to the $c$-axis is predicted to be the ground state, being favored by 0.117\,meV/Fe relative to placing the spins along the $a$-axis and 0.037\,meV/Fe relative to spins along the $b$-axis.  These anisotropies are in the typical range for non-cubic, $3d$ materials whose only magnetic elements are $3d$ elements; hcp Cobalt, for example exhibits a magnetic anisotropy of approximately 0.06\,meV/Co \cite{daalderop}.

\section*{Acknowledgments}

This work was supported by the U. S. Department of Energy, Office of Science, Basic Energy Sciences, Materials Sciences and Engineering Division. Work at ORNL’s High Flux Isotope reactor was supported by the Scientific User Facilities Division, Office of Basic Energy Sciences, U.S. Department of Energy (DOE).  We thank Huibo Cao for assistance with preliminary neutron diffraction experiments and Jiaqiang Yan for useful discussions.


\section*{Author contributions statement}

A.F.M. made the samples, performed physical property measurements and analysis, and wrote the paper; S.C. performed neutron diffraction and associated refinements; D.P. performed first principles calculations; B.C.S. motivated the study and discussed findings, M.A.M. performed diffraction experiments and took part in all critical discussions.  All authors reviewed the manuscript. 

\section*{Additional information}

There are no competing financial interests associated with this work. 


\renewcommand{\theequation}{S\arabic{equation}}
\renewcommand{\thefigure}{S\arabic{figure}}
\renewcommand{\thetable}{S\arabic{table}}

\setcounter{figure}{0}
\newpage
\section*{Supplementary Information}

\begin{center}
This Supplementary Information covers sample synthesis and thermal stability of CuFe$_2$Ge$_2$.  Electrical and thermal transport data are also reported, along with additional results from first principles calculations, and x-ray and neutron diffraction data and analysis.  Magnetization and x-ray diffraction data for the cobalt doped CuFeCoGe$_2$ sample are included as well.
\end{center}

\section*{Synthesis and Phase Stability}

The synthesis of CuFe$_2$Ge$_2$ can be achieved by arc-melting and annealing, and this study demonstrates that the annealing temperature utilized influences the observed properties.  Specifically, we find that samples annealed at 600\,$^{\circ}$C have an x-ray diffraction pattern that is well described by the reported structure and does not contain any significant impurities. This is consistent with the original report by Zavalij\cite{Zavalij1987}. A single impurity peak can sometimes be observed in the laboratory x-ray diffraction data, and can be indexed to Cu$_5$Ge$_2$.  Samples annealed at higher temperatures do not appear to be phase pure or of the same quality, which is here linked to disorder and/or decomposition.

\begin{figure}[ht]
\centering
\includegraphics[width=\linewidth]{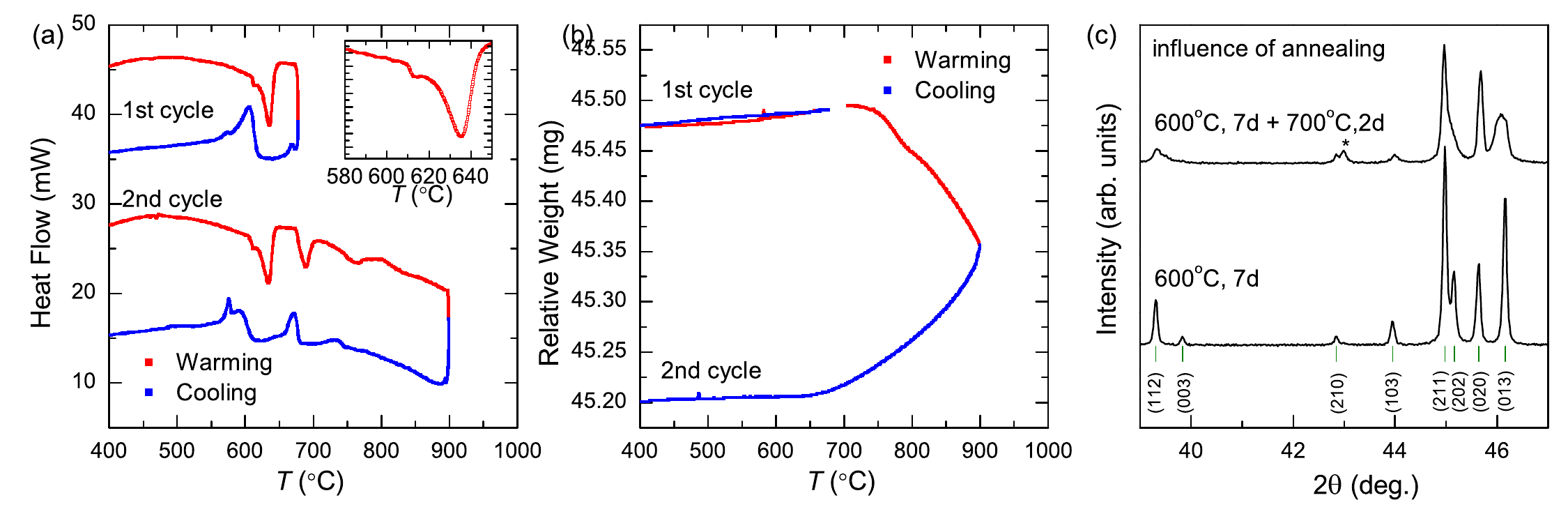}%
\caption{Thermogravimetric analysis and x-ray diffraction showing phase decomposition and/or disordering above $\approx\,635^{\circ}$C.}
\label{Decomp1}
\end{figure}

Differential thermal analysis (Fig.\,\ref{Decomp1}) revealed thermal anomalies at 612, 635, 685, and 750\,$^{\circ}$C, which were present upon cooling with a clear hysteresis and a little broadening.  The thermal anomalies remained upon heating the sample a second time.  Heating the $\approx$45\,mg sample above $\approx$750\,$^{\circ}$C produced measurable weight loss, likely due to formation of a volatile oxide upon heating.  Measurements were performed under flowing argon.

The events associated with the thermal anomalies shown in Fig.\,\ref{Decomp1} clearly affect the sample quality.  The influence of annealing at higher temperatures is shown in Fig.\,\ref{Decomp1}c, where powder x-ray diffraction data are shown in a characteristic region (common impurities Cu$_5$Ge$_2$, Fe$_{2-x}$Ge, Ge could all be observed in this region) .  A sample was formed in the reported way with a final step of annealing at 600\,$^{\circ}$C for 7\,d, and this sample has well-defined Bragg peaks that are easily indexed to the known structure.  A portion of this sample was then annealed for 2\,d at 700\,$^{\circ}$C.  After annealing at 700\,$^{\circ}$C, the diffraction peaks with an L component (H,K,L Miller indexes) are greatly broadened and an impurity peak near 43\,degrees 2$\Theta$ becomes evident (most likely Cu$_5$Ge$_2$).  The broadening of certain diffraction peaks would suggest disordering occurs, though the emergence of another peak clearly suggests a chemical change or decomposition.  On the contrary, the observation of thermal anomalies upon cooling (Fig.\,\ref{Decomp1}a) would suggest a reversible transition occurs, such as disordering or dissolution.

These results for phase stability have implications regarding crystal growth.   Crystals with magnetic properties similar to those of the polycrystalline samples annealed at 700\,$^{\circ}$C can be obtained using a self-flux (excess Cu and Ge).  However, these crystals grow at temperatures above the `decomposition' temperature, and thus it is not clear if they are stoichiometric at the composition CuFe$_2$Ge$_2$.  This is further complicated by the fact that the growth occurs in an environment with excess Cu and Ge.  Additional annealing and growth studies are required to fully understand the nature of the obtained crystals.  Searching for an inert (molten-metal) flux may be a better route to grow CuFe$_2$Ge$_2$ crystals below $\approx$630\,$^{\circ}$C.

\section*{Physical Property Data}

\begin{figure}[ht]
\centering
\includegraphics[width=\linewidth]{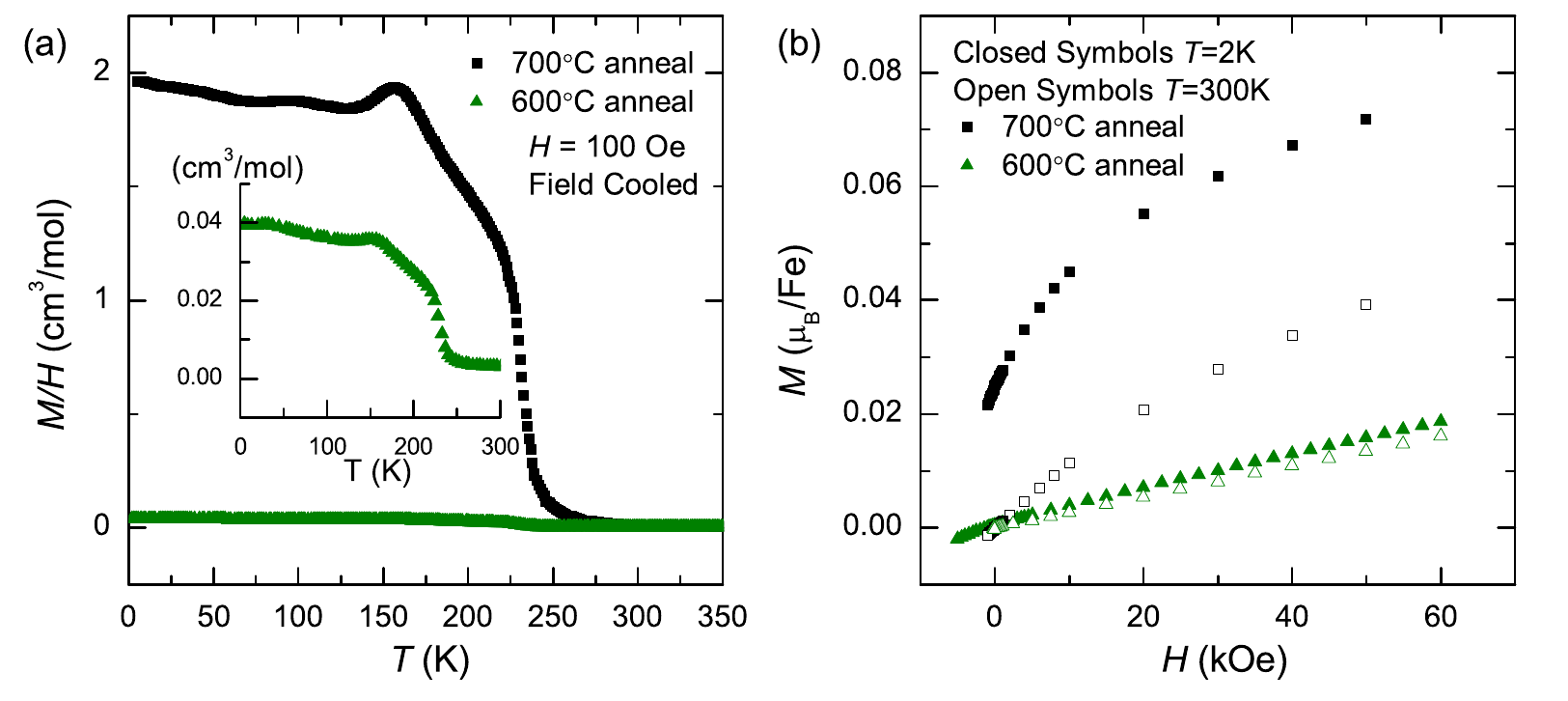}%
\caption{ Influence of annealing temperatures on magnetization data; legend labels 600\,$^{\circ}$C and 700\,$^{\circ}$C refer to sample annealing conditions as shown in Figure \ref{Decomp1}. After annealing at 700\,$^{\circ}$C the induced magnetization increases, as does the remanent magnetization observed at low $T$. The inset of (a) contains the same data as in the main panel on a scale that allows the temperature-dependence of $M$ to be observed.}
\label{Decomp2}
\end{figure}

Magnetization measurements were performed to further probe the changes that occur upon annealing at 700\,$^{\circ}$C (See Fig.\,\ref{Decomp2}).  These results clearly show an increase in the induced magnetization following annealing at 700\,$^{\circ}$C.  Interestingly, the temperature dependence of the induced magnetization $M$ remains qualitatively similar to that observed for the higher-quality samples produced by annealing at 600\,$^{\circ}$C.  As shown in Fig.\,\ref{Decomp2}b, the increased magnetization is present as an increase in the ferromagnetic contribution.  That is, the remanent moment $M_r$ is significantly larger in the sample that was annealed at 700\,$^{\circ}$C (at 2\,K $M_r$=0.025$\mu_B$/Fe versus 0.00046$\mu_B$/Fe for annealing at 600\,$^{\circ}$C).  One other way to compare the ferromagnetic contribution is to perform a linear fit to the high field data and compare the $M$ intercept.  Fitting between 3 and 5\,T yields zero-field (intercept) moments of 0.047$\mu_B$/Fe and 0.0016$\mu_B$/Fe at 2\,K for the samples annealed at 700 and 600\,$^{\circ}$C, respectively.   Given these results, it is tempting to conclude that the intrinsic behavior of crystalline CuFe$_2$Ge$_2$ is an antiferromagnetic ground state with no ferromagnetic contribution (zero remanent moment, zero intercept of the $M-H$ curve from high field).  However, the potential for canting of moments in this relatively complex magnetic structure cannot be ignored, nor can the potential of the field itself to modify the magnetic structure.  Regardless, though, it is clear that high quality CuFe$_2$Ge$_2$ should have a nearly linear $M(H)$ curve at low $T$ and display only a very small ferromagnetic contribution.

\begin{figure}[ht]
\centering
\includegraphics[width=\linewidth]{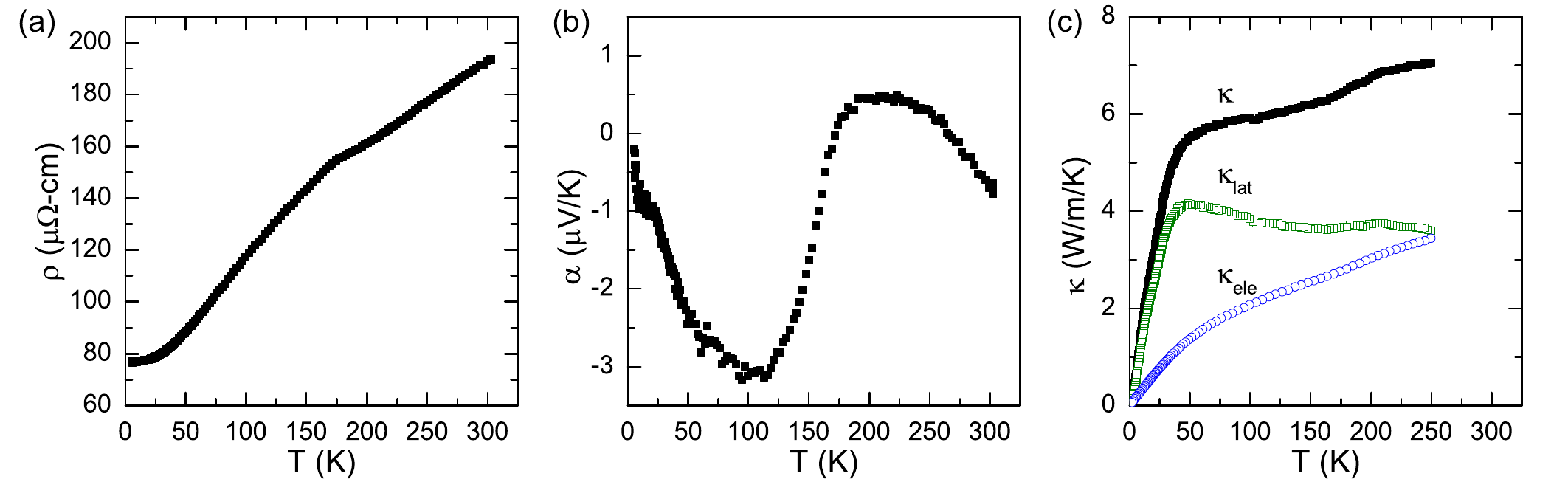}%
\caption{(a) Electrical resistivity, (b) Seebeck coefficient and (c) thermal conductivity of polycrystalline (hot-pressed) CuFe$_2$Ge$_2$.  In (c), the electronic contribution has been subtracted from the total thermal conductivity $\kappa$ to yield an estimate for the lattice contribution $\kappa_{lat}$.}
\label{TTO}
\end{figure}

To produce a dense compact for transport measurements, an arc-melted ingot was ground and hot-pressed at approximately 585$^{\circ}$C followed by annealing at 600$^{\circ}$C.  Hot-pressing occurred in a graphite furnace using a graphite die ($\frac{3}{8}$'' diameter) with an applied force of 500\,kg; the temperature was measured using a thermocouple inserted into the body of the die.  This yielded a pellet of approximately 87\% of the theoretical density, which is not as dense as desired for some transport measurements but is sufficient to provide insight into the connection between magnetism and transport.  In particular, the density should have little influence on the behavior of the Seebeck coefficient, which is not influenced by the magnitude of the electron relaxation time but rather by its energy dependence.  Thermal and electrical transport measurements were performed using the Thermal Transport Option of the Quantum Design PPMS.  Gold-coated copper leads were attached using silver epoxy.

Electrical and thermal transport measurements were performed to examine how the itinerant magnetism influenced the other physical properties, and the results are shown in Fig.\,\ref{TTO}.  The electrical resistivity displays a small change in slope near $T_N\approx$175\,K, but does not show any significant response at the onset of incommensurate spin density wave near 100-125\,K or at the onset of the enhanced magnetization near 230\,K.  The change in slope at $T_N$ is commonly observed across magnetic transitions, and likely indicates a reduced scattering of charge carriers below $T_N$ due to the loss of scattering from spin fluctuations.

The Seebeck coefficient ($\alpha$) responds to all three magnetic ordering temperatures, as shown in Fig.\,\ref{TTO}b.  A slope change is observed near $T_C \approx$ 230\,K, followed by a strong change at $T_N$ that leads to a sign change, and finally an extremum is observed at $\approx$100\,K.  The Seebeck coefficient is influenced by the shape of the Fermi surface and the scattering mechanisms of the free carriers.  As such, it can be a sensitive probe of the electrical transport. In the simplest model, a positive Seebeck coefficient indicates that holes dominate conduction while a negative Seebeck coefficient indicates electrons dominate conduction.  Thus, the change in sign of $\alpha$ is likely associated with a redistribution of the electronic states upon transitions between various magnetic states, and also indicates that multiple bands contribute to the Fermi surface.

The thermal conductivity is shown in Fig.\,\ref{TTO}c.  Also shown are estimates for the lattice contribution $\kappa_{lat}=\kappa-\kappa_{ele}$ and the electronic contribution obtained using the Wiedemann-Franz law, $\kappa_{ele}$=$L T/\rho$; the degenerate (metallic) limit of the Lorenz number $L$=2.44 $\times$ 10$^{-8}$\,W/$\Omega$/K$^2$ was utilized.  The data in Fig.\,\ref{TTO}c are limited to 250\,K, because radiation effects can cause erroneous data at higher temperatures.

Above $\approx$\,50\,K, the thermal conductivity increases with increasing temperature due to the increase in $\kappa_{ele}$.  In this same temperature range, the lattice contribution only decreases slightly.  This behavior deviates from the $1/T$ decay expected for crystalline lattice where phonon-phonon interactions dominate at high $T$.  As such, this suggests that some other scattering mechanisms are suppressing $\kappa$.  Given that $\kappa_{lat}$ appears to change slightly near $T_N$, it may be reasonable to suggest that the phonons are scattered by magnetic excitation and/or fluctuations.  Indeed, an interaction between thermal transport and magnetism is one additional manifestation of magnetoelastic coupling, consistent with the other observations of lattice-magnetism interactions.  The relatively high porosity of the polycrystalline compact may also produce excess scattering that reduces $\kappa_{lat}$ at all temperatures, though this would not lead to the small changes in $\kappa$ near 170-200\,K.

\section*{pXRD and Magnetism of CuFeCoGe$_2$}

\begin{figure}[ht]
\centering
\includegraphics[width=\linewidth]{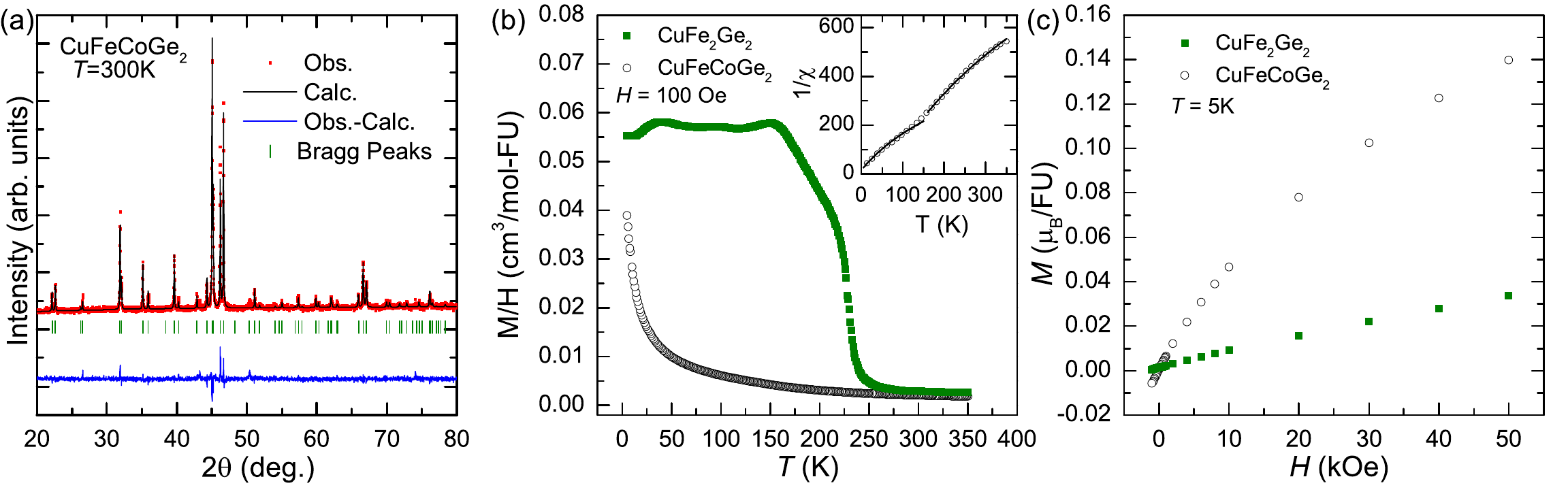}%
\caption{(a) Powder x-ray diffraction data for CuFeCoGe$_2$. (b) Temperature-dependence of the magnetization of CuFeCoGe$_2$ in comparison to CuFe$_2$Ge$_2$, with inset showing 1/$\chi$ data for CuFeCoGe$_2$. (c) Isothermal magnetization of CuFeCoGe$_2$ and CuFe$_2$Ge$_2$ at 5\,K.}
\label{cobalt}
\end{figure}

Cobalt doping was found to suppress the magnetism in CuFe$_2$Ge$_2$, allowing a comparison of the lattice properties in the magnetic CuFe$_2$Ge$_2$ to those in a non-magnetically-ordered material with similar chemical composition and structure.  A sample of nominal composition CuFeCoGe$_2$ was arc-melted at least five times and the resulting ingot was annealed at 600$^{\circ}$C for 7\,d; a piece of CuFe$_2$Ge$_2$ was made in the same way for a direct comparison.  The original structure paper for CuFe$_2$Ge$_2$ reported a cobalt-based phase of the same structure type, though partial mixing of the Cu and Co sites were reported\cite{Zavalij1987}.  Powder x-ray diffraction data for CuFeCoGe$_2$ at 300\,K are shown in Fig.\,\ref{cobalt}a, along with a Rietveld refinement in the expected structure; these data show phase purity equivalent to that in the undoped CuFe$_2$Ge$_2$ samples.

Magnetization data for the CuFeCoGe$_2$ sample is compared to that of CuFe$_2$Ge$_2$ in Fig.\,\ref{cobalt}b,c.  The ferromagnetic-like onset of the magnetization near 228\,K and the antiferromagnetic order are all suppressed by this nominal cobalt doping of 50\%.  When examining the behavior of 1/$\chi$, where the susceptibility $\chi$=$M$/$H$, there is a slight kink in the data for  CuFeCoGe$_2$ near 157(2)\,K.  This subtle feature can also be observed in the derivative d$\chi$/d$T$ (not shown).  The exact nature of this anomaly is not known. The isothermal magnetization is shown in Fig.\,\ref{cobalt}c, which shows the cobalt-containing sample to be more easily polarized compared to the antiferromagnetically-ordered CuFe$_2$Ge$_2$.  This may suggest that finer control of the cobalt doping could lead to a ferromagnetic ground state, though a detailed study into the physical properties of cobalt-doped samples should be performed before any conclusions are drawn. In any case, CuFeCoGe$_2$ appears to be a suitable material to utilize as a non-magnetically-ordered analogue when considering the evolution of the lattice parameters as a function of $T$ in CuFe$_2$Ge$_2$.

\section*{First principles calculations}

\begin{figure}[ht]
\centering
\includegraphics[width=00.5\linewidth]{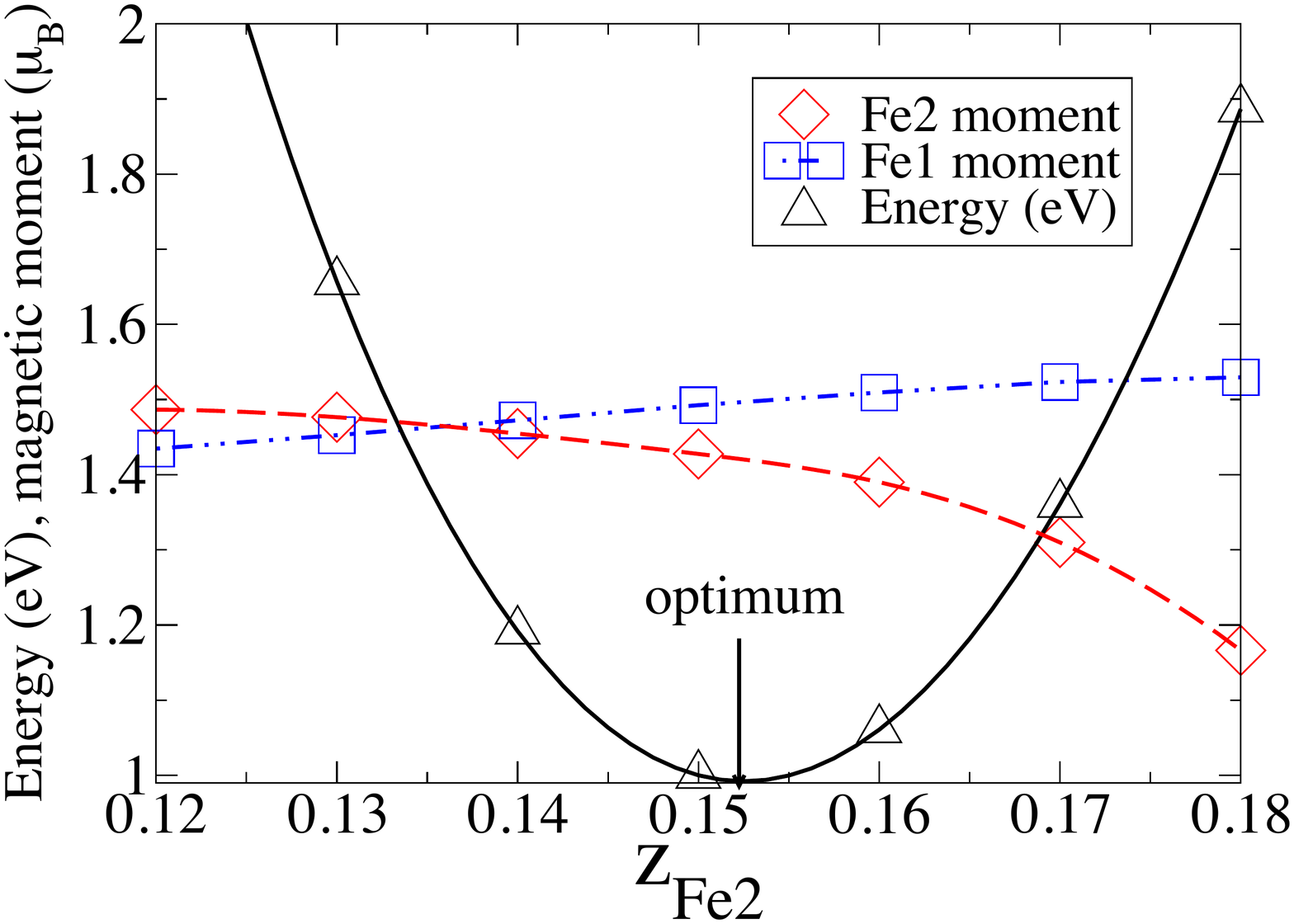}%
\caption{Calculation probing the influence of the Fe(2) $z$-coordinate on the Fe(1) and Fe(2) moments, as well as the total energy within the calculated AF0 ground state.}
\label{zPlot}
\end{figure}

First principles calculations were utilized to explore the magnetoelastic coupling and to aid in the refinement of neutron diffraction data.  In general, our calculation results are in agreement with those presented by Shanavas and Singh\cite{Shanavas2015}.  To provide a more direct comparison with the Fe-based superconductor systems, which also display significant magnetoelastic coupling, the magnetic moments were calculated as a function of the Fe(2) $z$ coordinate.  An increase in this $z$ represents a reduction in the Fe(1)-Fe(2) bond distance, and a further spacing of the Fe(2)-Fe(2) distance that separates the sawtooth chains of Fe.  The calculation results are shown in Fig.\,\ref{zPlot}, and opposite trends for the two Fe sites are observed.  In general, though, the change is not as significant as that observed in the Fe-based superconducting materials.  These calculations were done within the AF0 ground state, as described in the Methods Section of the main text.

The coupling between the imposed magnetic structure and the crystal structure was also investigated with first principles calculations.  The unconstrained atomic coordinates ($z$ of Ge(1), Ge(2), and Fe(2)), were relaxed after imposing a given magnetic structure.  The results are summarized in Table\,\ref{DFT}.  This calculation revealed that the ferromagnetic configuration, which we obtain to have the largest moments, produces the greatest response in the lattice.  This calculation also reveals the proximity of the various magnetic ground states.

\begin{table}[h!]
\caption{\label{DFT}Calculated results for the non-magnetically-ordered state and several magnetic states.}
\begin{center}
\begin{tabular}{|c|c|c|c|}
\hline
State & z$_{Fe_{2}}$ & $\Delta$ E  (meV/Fe) & Staggered moment Fe(1), Fe(2)\\ \hline
Non-magnetic & 00.1541 & - &- \\ \hline
Ferromagnetic & 00.1636 & -58 &  +10.55, +10.50\\ \hline
AF1 & 00.1524 & -84 &  $\pm$ 1.32, $\pm$ 1.32\\ \hline
AF0 & 00.1521 & -99 & $\pm$ 1.49, $\pm$ 1.41\\ \hline
\end{tabular}
\end{center}
\end{table}

\newpage

\section*{X-ray and Neutron Diffraction Data and Analysis}

\begin{figure}[ht]
\centering
\includegraphics[width=\linewidth]{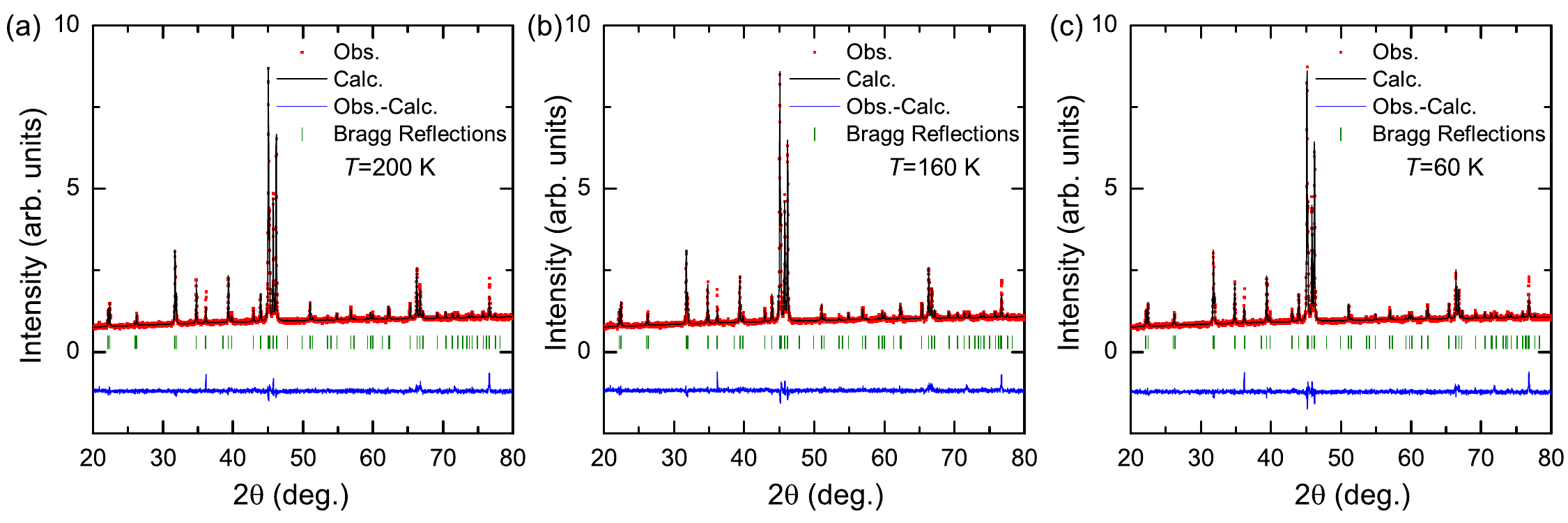}%
\caption{X-ray diffraction data and Rietveld refinements at (a) 200\,K, (b) 160\,K, and (c) 60\,K.}
\label{pxrd3T}
\end{figure}

To complement the temperature-dependent neutron diffraction results, powder x-ray diffraction data at $T$=200, 160, and 60\,K are presented in Figure \ref{pxrd3T}.  These results demonstrate that the nuclear structure does not fundamentally change across the magnetic transitions.  As noted in the main text, however, the $c$ lattice parameter clearly responds to the magnetic transitions.

\begin{figure}[ht]
\centering
\includegraphics[width=\linewidth]{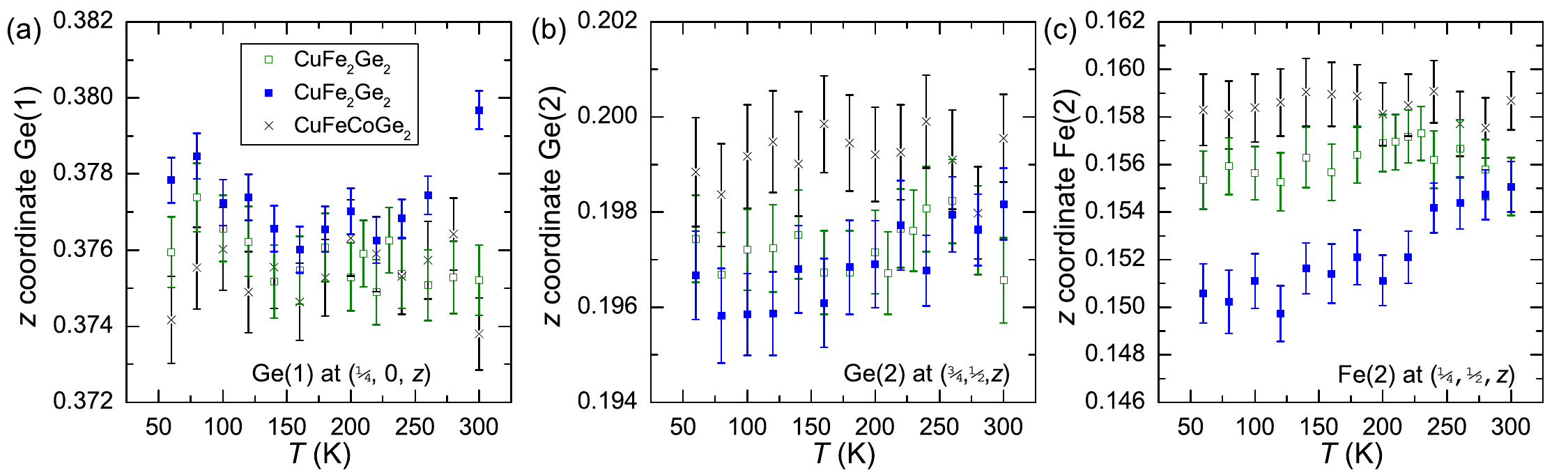}%
\caption{Atomic coordinates from Rietveld refinements of x-ray diffraction data for CuFe$_2$Ge$_2$ and CuFeCoGe$_2$.  As in the main text, data for two different CuFe$_2$Ge$_2$ samples are shown (the open symbols correspond to the sample used for neutron diffraction).}
\label{coords}
\end{figure}

\begin{figure}[ht!]
\centering
\includegraphics[width=\linewidth]{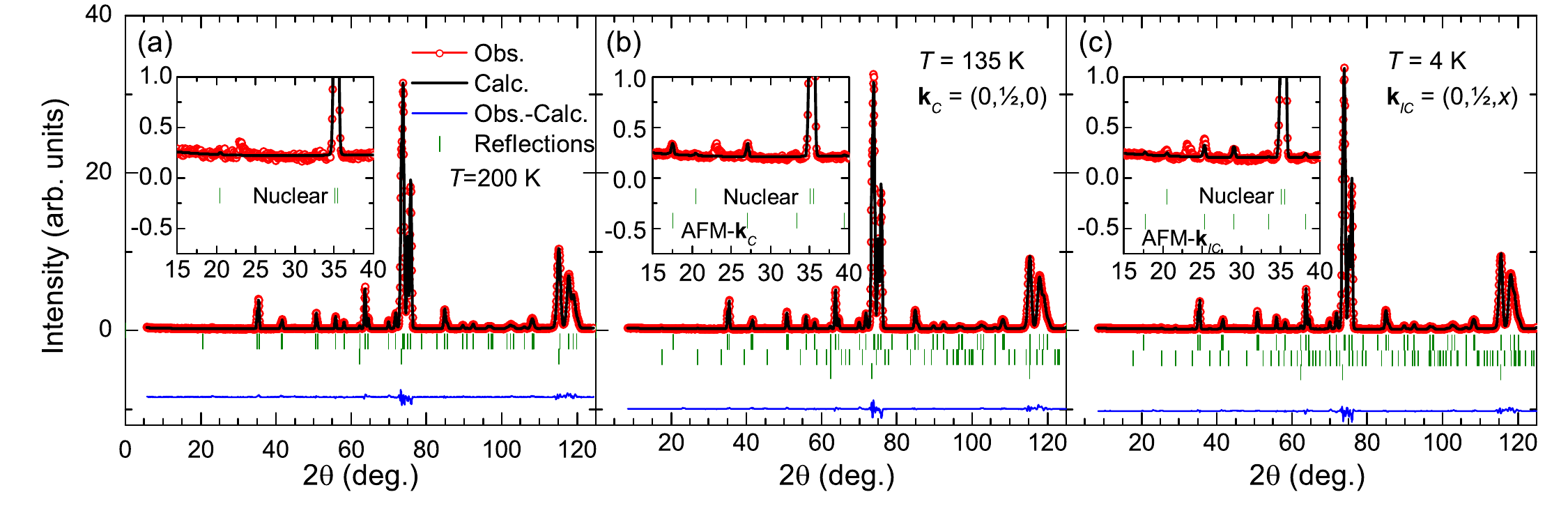}%
\caption{Neutron diffraction data and Rietveld refinements at (a) 200\,K, (b) 135\,K, and (c) 4\,K.  The inset highlights the region where magnetic scattering is indexed, with propagation vectors \textbf{k}$_{C}$ and \textbf{k}$_{IC}$ provided in the legend.}
\label{HB2A}
\end{figure}

Figure \ref{coords} shows the refined values of the unconstrained atomic coordinates for CuFe$_2$Ge$_2$ obtained from powder x-ray diffraction data.  The corresponding lattice parameters are shown in Figure 4 of the main text.  Data for two CuFe$_2$Ge$_2$ samples are shown, as is also shown in Figure 4 of the main text, along with data obtained for the sample of nominal composition CuFeCoGe$_2$.  As observed in Fig.\,\ref{coords}, the atomic coordinates are relatively independent of temperature and fairly similar for the different samples.  Higher quality diffraction data, ideally from single crystals, would be required to better isolate any small shifts in atomic positions that occur in relation to the magnetic transitions.

Complete neutron diffraction datasets were collected at several temperatures, and data are shown for three pertinent temperatures in Figure \ref{HB2A}.  Data collected at $T=200\,K$, which is below the ferromagnetic-like onset at 228\,K, were well-described using the non-magnetic structure utilized for refinement of x-ray diffraction data.  The data at $T=135$\,K were refined using the magnetic propagation vector \textbf{k}$_{C}$=(0,$\frac{1}{2}$,0), and the inset of Figure \ref{HB2A}b highlights the contribution of magnetic scattering.  The magnetic intensity of this commensurate \textit{AFM-C} phase was greater at 135\,K than any other temperature examined in detail.  The lower $T$, incommensurate structure \textit{AFM-IC} with propagation vector \textbf{k}$_{IC}$=(0,$\frac{1}{2}$,$x$) was utilized to model the data collected at 4\,K.  The non-indexed peak near 22 degrees 2$\theta$ in Figure \ref{HB2A} likely comes from alloys in the Al can, though it could also be due to a minor impurity within the sample.

\newpage
  \textit{}
\newpage
\subsection*{Representational Analysis $T$=135\,K}


\begin{table}[ht]
\caption{Basis vectors for the space group $P m m a$ with 
\textbf{k}$_{C}=( 0,~ 0.5,~ 0)$.The 
decomposition of the magnetic representation for 
the $Fe$ site 
$( 0,~ 0.5,~ 0.5)$ is 
$\Gamma_{Mag}=0\Gamma_{1}^{1}+1\Gamma_{2}^{1}+0\Gamma_{3}^{1}+2\Gamma_{4}^{1}+0\Gamma_{5}^{1}+1\Gamma_{6}^{1}+0\Gamma_{7}^{1}+2\Gamma_{8}^{1}$. The atoms of the nonprimitive 
basis are defined according to 
1: $( 0,~ 0.5,~ 0.5)$, 2: $( 0.5,~ 0.5,~ 0.5)$.}
\label{basis_vector_table_1}
\begin{tabular}{ccc|cccccc}
\hline
  IR  &  BV  &  Atom & \multicolumn{6}{c}{BV components}\\
      &      &             &$m_{\|a}$ & $m_{\|b}$ & $m_{\|c}$ &$im_{\|a}$ & $im_{\|b}$ & $im_{\|c}$ \\
\hline
$\Gamma_{2}$ & $\bfpsi_{1}$ &      1 &      0 &      4 &      0 &      0 &      0 &      0  \\
             &              &      2 &      0 &      4 &      0 &      0 &      0 &      0  \\
$\Gamma_{4}$ & $\bfpsi_{2}$ &      1 &      4 &      0 &      0 &      0 &      0 &      0  \\
             &              &      2 &      4 &      0 &      0 &      0 &      0 &      0  \\
             & $\bfpsi_{3}$ &      1 &      0 &      0 &      4 &      0 &      0 &      0  \\
             &              &      2 &      0 &      0 &     -4 &      0 &      0 &      0  \\
$\Gamma_{6}$ & $\bfpsi_{4}$ &      1 &      0 &      4 &      0 &      0 &      0 &      0  \\
             &              &      2 &      0 &     -4 &      0 &      0 &      0 &      0  \\
$\Gamma_{8}$ & $\bfpsi_{5}$ &      1 &      4 &      0 &      0 &      0 &      0 &      0  \\
             &              &      2 &     -4 &      0 &      0 &      0 &      0 &      0  \\
             & $\bfpsi_{6}$ &      1 &      0 &      0 &      4 &      0 &      0 &      0  \\
             &              &      2 &      0 &      0 &      4 &      0 &      0 &      0  \\
\hline
\end{tabular}
\end{table}

\begin{table}[ht]
\caption{Basis vectors for the space group $P m m a$ with 
\textbf{k}$_{C}=( 0,~ 0.5,~ 0)$.The 
decomposition of the magnetic representation for 
the $Fe$ site 
$( .25,~ 0.5,~ 0.15507)$ is 
$\Gamma_{Mag}=1\Gamma_{1}^{1}+1\Gamma_{2}^{1}+1\Gamma_{3}^{1}+1\Gamma_{4}^{1}+1\Gamma_{5}^{1}+0\Gamma_{6}^{1}+0\Gamma_{7}^{1}+1\Gamma_{8}^{1}$. The atoms of the nonprimitive 
basis are defined according to 
1: $(0.25,~ 0.5,~ 0.15507)$, 2: $( 0.75,~ 0.5,~ 0.84493)$.}
\label{basis_vector_table_2}
\begin{tabular}{ccc|cccccc}
\hline
  IR  &  BV  &  Atom & \multicolumn{6}{c}{BV components}\\
      &      &             &$m_{\|a}$ & $m_{\|b}$ & $m_{\|c}$ &$im_{\|a}$ & $im_{\|b}$ & $im_{\|c}$ \\
\hline
$\Gamma_{1}$ & $\bfpsi_{1}$ &      1 &      4 &      0 &      0 &      0 &      0 &      0  \\
             &              &      2 &     -4 &      0 &      0 &      0 &      0 &      0  \\
$\Gamma_{2}$ & $\bfpsi_{2}$ &      1 &      0 &      4 &      0 &      0 &      0 &      0  \\
             &              &      2 &      0 &      4 &      0 &      0 &      0 &      0  \\
$\Gamma_{3}$ & $\bfpsi_{3}$ &      1 &      0 &      4 &      0 &      0 &      0 &      0  \\
             &              &      2 &      0 &     -4 &      0 &      0 &      0 &      0  \\
$\Gamma_{4}$ & $\bfpsi_{4}$ &      1 &      4 &      0 &      0 &      0 &      0 &      0  \\
             &              &      2 &      4 &      0 &      0 &      0 &      0 &      0  \\
$\Gamma_{5}$ & $\bfpsi_{5}$ &      1 &      0 &      0 &      4 &      0 &      0 &      0  \\
             &              &      2 &      0 &      0 &     -4 &      0 &      0 &      0  \\
$\Gamma_{8}$ & $\bfpsi_{6}$ &      1 &      0 &      0 &      4 &      0 &      0 &      0  \\
             &              &      2 &      0 &      0 &      4 &      0 &      0 &      0  \\
\hline
\end{tabular}
\end{table}

\newpage

\subsection*{Representational Analysis $T$=4\,K}

\begin{table}[ht]
\caption{Basis vectors for the space group $P m m a$ with 
\textbf{k}$_{IC}=( 0,~ 0.5,~ 0.125)$.The 
decomposition of the magnetic representation for 
the $Fe$ site 
$( 0,~ 0.5,~ 0.5)$ is 
$\Gamma_{Mag}=2\Gamma_{1}^{1}+1\Gamma_{2}^{1}+1\Gamma_{3}^{1}+2\Gamma_{4}^{1}$. The atoms of the nonprimitive 
basis are defined according to 
1: $( 0,~ 0.5,~ 0.5)$, 2: $( 0.5,~ 0.5,~ 0.5)$.}
\label{basis_vector_table_1b}
\begin{tabular}{ccc|cccccc}
\hline
  IR  &  BV  &  Atom & \multicolumn{6}{c}{BV components}\\
      &      &             &$m_{\|a}$ & $m_{\|b}$ & $m_{\|c}$ &$im_{\|a}$ & $im_{\|b}$ & $im_{\|c}$ \\
\hline
$\Gamma_{1}$ & $\bfpsi_{1}$ &      1 &      2 &      0 &      0 &      0 &      0 &      0  \\
             &              &      2 &      2 &      0 &      0 &      0 &      0 &      0  \\
             & $\bfpsi_{2}$ &      1 &      0 &      0 &      2 &      0 &      0 &      0  \\
             &              &      2 &      0 &      0 &     -2 &      0 &      0 &      0  \\
$\Gamma_{2}$ & $\bfpsi_{3}$ &      1 &      0 &      2 &      0 &      0 &      0 &      0  \\
             &              &      2 &      0 &      2 &      0 &      0 &      0 &      0  \\
$\Gamma_{3}$ & $\bfpsi_{4}$ &      1 &      0 &      2 &      0 &      0 &      0 &      0  \\
             &              &      2 &      0 &     -2 &      0 &      0 &      0 &      0  \\
$\Gamma_{4}$ & $\bfpsi_{5}$ &      1 &      2 &      0 &      0 &      0 &      0 &      0  \\
             &              &      2 &     -2 &      0 &      0 &      0 &      0 &      0  \\
             & $\bfpsi_{6}$ &      1 &      0 &      0 &      2 &      0 &      0 &      0  \\
             &              &      2 &      0 &      0 &      2 &      0 &      0 &      0  \\
\hline
\end{tabular}
\end{table}

\begin{table}[ht]
\caption{Basis vectors for the space group $P m m a$ with 
\textbf{k}$_{IC}=( 0,~ 0.5,~ 0.125)$.The 
decomposition of the magnetic representation for 
the $Fe$ site 
$( .25,~ 0.5,~ 0.15533)$ is 
$\Gamma_{Mag}=1\Gamma_{1}^{1}+1\Gamma_{2}^{1}+0\Gamma_{3}^{1}+1\Gamma_{4}^{1}$. The atom of the primitive 
basis is defined according to 
1: $( .25,~ 0.5,~ 0.15533)$.}
\label{basis_vector_table_2b}
\begin{tabular}{ccc|cccccc}
\hline
  IR  &  BV  &  Atom & \multicolumn{6}{c}{BV components}\\
      &      &             &$m_{\|a}$ & $m_{\|b}$ & $m_{\|c}$ &$im_{\|a}$ & $im_{\|b}$ & $im_{\|c}$ \\
\hline
$\Gamma_{1}$ & $\bfpsi_{1}$ &      1 &      4 &      0 &      0 &      0 &      0 &      0  \\
$\Gamma_{2}$ & $\bfpsi_{2}$ &      1 &      0 &      4 &      0 &      0 &      0 &      0  \\
$\Gamma_{4}$ & $\bfpsi_{3}$ &      1 &      0 &      0 &      4 &      0 &      0 &      0  \\
\hline
\end{tabular}
\end{table}

\begin{table}[ht]
\caption{Basis vectors for the space group $P m m a$ with 
\textbf{k}$_{IC}=( 0,~ 0.5,~ 0.125)$.The 
decomposition of the magnetic representation for 
the $Fe_2$ site 
$( 0.75,~ 0.5,~ 0.84467)$ is 
$\Gamma_{Mag}=1\Gamma_{1}^{1}+1\Gamma_{2}^{1}+0\Gamma_{3}^{1}+1\Gamma_{4}^{1}$. The atom of the primitive 
basis is defined according to 
1: $( 0.75,~ 0.5,~ 0.84467)$.}
\label{basis_vector_table_3}
\begin{tabular}{ccc|cccccc}
\hline
  IR  &  BV  &  Atom & \multicolumn{6}{c}{BV components}\\
      &      &             &$m_{\|a}$ & $m_{\|b}$ & $m_{\|c}$ &$im_{\|a}$ & $im_{\|b}$ & $im_{\|c}$ \\
\hline
$\Gamma_{1}$ & $\bfpsi_{1}$ &      1 &      4 &      0 &      0 &      0 &      0 &      0  \\
$\Gamma_{2}$ & $\bfpsi_{2}$ &      1 &      0 &      4 &      0 &      0 &      0 &      0  \\
$\Gamma_{4}$ & $\bfpsi_{3}$ &      1 &      0 &      0 &      4 &      0 &      0 &      0  \\
\hline
\end{tabular}
\end{table}

\end{document}